\documentclass[12pt]{article}
\usepackage{amsfonts,amsmath}
\usepackage{color}
\usepackage{tabularx}

\usepackage{times}

\usepackage[dvips]{graphicx}
\usepackage{ifthen}
\usepackage{enumerate}
\usepackage{amsmath}
\usepackage{amssymb}
\usepackage[mathscr]{eucal}
\usepackage[errorshow]{tracefnt}
\usepackage{amsmath}
\usepackage{amssymb}
\usepackage{array}
\usepackage{amsthm}
\usepackage{eucal}
\usepackage{eufrak}
\usepackage{epsf}
\usepackage{epsfig}
\usepackage{psfrag}
\usepackage{multirow}
\usepackage[apple mac]{inputenc}

\def\+{{+\!\!\!+}}

\def\la{\label}

\def\Im           {{\rm Im\hskip0.1em}}

\def\pmb#1{\setbox0=\hbox{#1}%
\kern.0em\copy0\kern-\wd0
\kern-.04em\copy0\kern-\wd0
\kern.08em\copy0\kern-\wd0
\kern-.04em\raise.0433em\box0 }         

\newcommand{\nc}{\newcommand}
\nc{\beq}{\begin{equation}}
\nc{\eeq}[1]{\label{#1}\end{equation}}
\nc{\ber}{\begin{eqnarray}}
\nc{\eer}[1]{\label{#1}\end{eqnarray}}
\nc{\pek}[1]{\cite{#1}}
\nc{\enr}[1]{(\ref{#1})}
\nc{\kal}[1]{{\cal{#1}}}
\nc{\dott}{\;\cdot\;}

\def\0 {\nonumber}

\newcommand{\bea}{\begin{eqnarray}}
\newcommand{\eea}{\end{eqnarray}}

\newcommand{\CD}{{\mathcal D}}
\newcommand{\CF}{{\mathcal F}}

\newcommand{\CN}{{\mathcal N}}
\newcommand{\CO}{{\mathcal O}}

\renewcommand\Im{{\mathrm{Im}}}

\numberwithin{equation}{section}
\newcommand{\be}{\begin{equation}}
\newcommand{\ee}{\end{equation}}

\begin{document}
\begin{titlepage}
\begin{center}

\hfill SISSA  82/2010/EP-FM \\
\hfill IHES/P/10/43\\

\vskip .8in \noindent


{\LARGE \bf{ Generalized matrix models }}\\ 

\vskip .1in

{\LARGE \bf{ and }} \\

\vskip .1in

{\LARGE \bf{ AGT correspondence at all genera }} \\

\vskip .7in

{\bf Giulio Bonelli$^{\heartsuit}$, Kazunobu Maruyoshi$^{\heartsuit}$, Alessandro Tanzini$^{\heartsuit}$ and Futoshi Yagi$^{\clubsuit}$}

\vskip .2
in
{\em\small
$^{\heartsuit}$ International School of Advanced Studies (SISSA) \\via Bonomea 265, 34136 Trieste, Italy 
and INFN, Sezione di Trieste \\
$^{\clubsuit}$ Institut des Hautes \'Etudes Scientifiques, 91440, Bures-sur-Yvette, France }

\vskip .6in
\end{center}
\begin{center} {\bf ABSTRACT }
\end{center}
\begin{quotation}\noindent

  We study generalized matrix models corresponding to $n$-point 
  Virasoro conformal blocks on Riemann surfaces with arbitrary genus $g$. Upon AGT correspondence,
  these  
  describe four dimensional $\CN=2$ $SU(2)^{n+3g-3}$ gauge theories 
  with generalized quiver diagrams.
  We obtain the generalized matrix models from the
  perturbative evaluation of the Liouville correlation functions and verify the consistency of the description
  with respect to degenerations of the Riemann surface.
  Moreover, we derive the Seiberg-Witten curve for the $\CN=2$ gauge theory as the
  spectral curve of the generalized matrix model, thus providing a check of AGT correspondence at all genera. 

\end{quotation}
\vfill
\eject

\end{titlepage}
\tableofcontents

\section{Introduction}
\label{sec:intro}

The distinctive feature of M-theory is the description in geometrical terms of
non-perturbative phases of superstrings. This approach is very effective 
for local geometries, where the dynamics of gravitational 
degrees of freedom gets decoupled and we gain a framework for the description
of non-perturbative gauge theory dynamics. M-theory beautifully encodes
the Seiberg-Witten geometry of four dimensional ${\cal N}=2$ theories 
in terms of M5-brane compactifications \cite{Lerche,Witten,Gaiotto}.
In particular in \cite{Gaiotto} a full class of generalized quiver
gauge theories has been described in terms of multiple M5-brane systems
covering a generic punctured Riemann surface ${\cal C}_{g,n}$.
For example, for ${\cal C}_{0,n}$ and ${\cal C}_{1,n}$ one recovers Witten's constructions 
of linear and circular quivers in the appropriate degeneration limits.

In this context a very intriguing relation between the partition function of four dimensional $SU(2)^{n+3g-3}$ 
superconformal
${\cal N}=2$ gauge theories \cite{Nekrasov} and Liouville theory on ${\cal C}_{g,n}$ has been discovered in \cite{AGT}.
This proposal has been a subject of intensive investigations and refinements from different viewpoints. 
Evidence for this conjecture as well as complete proofs for some cases can be found in \cite{proof, Alba:2009ya}. 
Extensions to higher rank gauge groups and Toda field theories were introduced and discussed in \cite{sun}.
The refinement of the correspondence in presence of gauge theory observables 
has been presented and studied in \cite{AGGTV, surface, MT}.
Moreover, some arguments for the derivation of the AGT correspondence were proposed 
in the M-theory context in \cite{mtheory}
and via matrix models in \cite{DV, IMO, mm1,mm2, mm, EM, EM2, MY, MMMsurface, CDV}.

Here we would like to address this correspondence from a complementary point of view, explaining how to recover 
the geometry of the M-theory set-up and the Seiberg-Witten data in the wildest generality. 
To this end we derive a generalized matrix model from Liouville theory on ${\cal C}_{g,n}$ 
and study its large $N$ limit recovering the gauge theory Seiberg-Witten curve 
as its spectral curve.
This provides a check of the AGT conjecture at all genera.

In Section 2 we derive the generalized matrix model -- as extended Selberg integrals -- starting from the Coulomb gas 
representation of the residues of the perturbative Liouville theory correlators. 
The matrix model potential that we get has the form anticipated by \cite{DV} and 
in the elliptic case it coincides with the one derived in \cite{MY}.

In Section 3 we discuss the stability of this picture and its consistency with respect to the degeneration of the 
curve ${\cal C}_{g,n}$ in general and present the degenerations of punctured tori as an explicative example.

In Section 4 we analyze the large $N$ limit and show how, by consistently adapting to our case 
the standard matrix model techniques, one gets a spectral curve in terms of quadratic differentials on 
${\cal C}_{g,n}$ precisely reproducing the Seiberg-Witten curve and differential proposed in \cite{Gaiotto}.

We leave our concluding remarks to Section 5 and devote an Appendix to the detailed study of the degenerations of the 
${\cal C}_{2,0}$ Seiberg-Witten data.

\section{From Liouville theory to generalized matrix model}
\label{sec:Liouville}
  In this section we derive the generalized matrix model
  which corresponds to the $n$ point conformal block on a Riemann surface ${\cal C}_g$ of  genus $g$.
  We derive it from the perturbative calculation of the 
  correlation function of the Liouville theory
  by following the discussion in \cite{Goulian:1990qr}.

  The $n$-point function of the Liouville theory on ${\cal C}_g$ is  given by the following path integral 
    \bea
    A 
     \equiv \left< \prod_{k=1}^n e^{ - 2 m_k\phi(w_k, \bar{w}_k)} \right>_{{\rm Liouville \,\, on}\,\, {\cal C}_g}
     \equiv \int {\mathcal{D}} \phi (z,\bar{z}) e^{-S[\phi]} \prod_{k=1}^n e^{- 2m_k \phi (w_k,\bar{w}_k)},
    \eea
  where the Liouville action is given by 
    \bea
    S[\phi]
     =     \frac{1}{4 \pi} \int d^2 z \sqrt{g} 
           (g^{ab} \partial_a \phi \partial_b \phi + Q R \phi + 4 \pi \mu e^{2 b \phi}).
    \eea
  We divide the Liouville field into the zero mode and the fluctuation
  $\phi (z,\bar{z}) = \phi_0 + \tilde{\phi} (z,\bar{z})$, obtaining
    \bea
    A
     =     \int \CD \tilde{\phi} e^{- \tilde{S}} \prod_{k=1}^n e^{- 2m_k \tilde{\phi} (w_k, \bar{w}_k)}
           \int_{-\infty}^{+\infty} d \phi_0 e^{- \mu e^{2 b \phi_0} \int d^2 z \sqrt{g} e^{2 b \tilde{\phi}}}
           e^{- \frac{Q \phi_0}{4 \pi} \int d^2 z \sqrt{g} R } e^{-2 \phi_0 \sum_k m_k},
    \eea
  where 
    \bea
    \tilde{S}
     =     \frac{1}{4 \pi} \int d^2 z \sqrt{g} (g^{ab} \partial_a \tilde{\phi} \partial_b \tilde{\phi} + Q R \tilde{\phi}).
    \label{stilda}
    \eea
  We can integrate out the zero mode $\phi_0$ as
    \bea
    \int_{-\infty}^{+\infty} d \phi_0 \ e^{- \mu e^{2 b \phi_0} \int d^2 z \sqrt{g} e^{2 b \tilde{\phi}}}
           e^{- 2 (g - 1) Q \phi_0 } e^{-2 \phi_0 \sum_k m_k}
     =     \frac{\mu^N \Gamma(-N)}{2b} \left( \int d^2 z \sqrt{g} e^{2 b \tilde{\phi}} \right)^N,
    \eea
  where we have used $\int d^2 z \sqrt{g} R = 4 \pi \chi = 8 \pi (1-g)$
  and $N$ is defined as
    \bea
    N
     \equiv    \frac{1}{b} \sum_k m_k + \frac{Q}{b} (1-g).
           \label{mom_cons}
    \eea
  Therefore, the $n$-point function can be written as
    \bea
    A
     =     \frac{\mu^N \Gamma(-N)}{2b} \int \CD \tilde{\phi} e^{- \tilde{S}} 
           \left( \int d^2 z \sqrt{g} e^{2 b \tilde{\phi}} \right)^N
           \prod_{k=1}^n e^{-2m_k \tilde{\phi} (w_k, \bar{w}_k)} .
    \eea
  When $ N \in \mathbb{Z}_{\geq 0} $, 
  the correlator diverges due to the factor $\Gamma(-N)$. The residues $A_N$ at these simple poles are computed then
  in perturbation theory in $b$ around the free scalar field action (\ref{stilda}). 
  From now on, our convention is that
    \bea
    \left< \ldots \right>_{{\rm free ~on} ~\mathcal{C}_g}
     =     \int \CD \tilde{\phi} e^{- \frac{1}{4 \pi} 
           \int d^2 z \sqrt{g} g^{ab} \partial_a \tilde{\phi} \partial_b \tilde{\phi}} \ldots,
    \eea
  which leads to
    \bea
    A_N
     =     \frac{(-\mu)^N }{2b N!} \left< e^{- \frac{Q}{4 \pi} \int d^2 z \sqrt{g} R \tilde{\phi}(z)} 
           \prod_{i = 1}^N \int d^2 z_i \sqrt{g} e^{2 b \tilde{\phi}(z_i)}
           \prod_{k=1}^n e^{- 2 m_k \tilde{\phi}(w_k)} \right>_{{\rm free ~on} ~\mathcal{C}_g}.
           \label{A1}
    \eea
  The condition (\ref{mom_cons}) ensures momentum conservation in the free theory.
  
  Here we choose as a reference volume form $d^2 z \sqrt{g} = |\omega(z) dz|^2$
  where $\omega(z)$ is the coefficient of a reference holomorphic differential.
  This differential has $2 g - 2$ zeros, which we denote by $\xi_I$ ($I = 1, \cdots 2g-2$).
  Then, the first factor in the expectation value of (\ref{A1}) becomes
    \bea
    \frac{Q}{2 \pi} \int d^2 z \  \tilde{\phi}(z)  \partial \bar{\partial} {\log {|\omega|^2}}
     =     \frac{Q}{2 \pi} \int d^2 z \ \tilde{\phi}(z) \sum_{I=1}^{2g-2} (2 \pi) \delta^2 (z - \xi_I)
     =     Q \sum_{I=1}^{2g-2} \tilde{\phi} (\xi_I),
    \eea
  where we have used $R = - (2/\sqrt{g}) \partial \bar{\partial} {\log} \sqrt{g}$.
  Thus, we obtain
    \bea
    A_N
     =     \frac{(-\mu)^N }{2b N!} \left< \prod_{I=1}^{2g-2} e^{ Q \tilde{\phi} (\xi_I)}
           \int \prod_{i = 1}^N d^2 z_i |\omega(z_i)|^2 e^{2 b \tilde{\phi}(z_i)}
           \prod_{k=1}^n e^{- 2 m_k \tilde{\phi}(w_k)} \right>_{{\rm free ~on} ~\mathcal{C}_g} .
    \eea
  The $\ell$-point function of the free theory on ${\cal C}_g$ is given 
  in the factorized form as \cite{VV, DVV, dhp}
    \bea
    & &    \left< \prod_{i=1}^\ell e^{i k_i \phi(z_i, \bar{z}_i)} \right>_{{\rm free ~on} ~\mathcal{C}}
     =     \left({\det}{\Im}\tau\right)^{1/2}
           \delta(\sum_i k_i) \times \nonumber\\
    & &    \int_{- \infty}^{\infty} \prod_{a=1}^g d p_a
           \left| \prod_{i=1}^{\ell} \omega(z_i)^{-k_i^2/4} \prod_{i<j} E(z_i, z_j)^{k_i k_j/2} 
           \exp \left( 2 \pi i \sum_{a, b} p_a p_b \tau_{ab}
           + 2 \pi i \sum_{a, i} p_a k_i \int^{z_i} \omega_a \right) \right|^2,
           \nonumber \\
           \label{factorized} 
    \eea
  where $\tau_{ab}$ is the period matrix,
  $E \left( z_{i} , z_{j} \right)$ is the prime form,
  $\{ \omega_a\}$ is a basis of normalized holomorphic one-forms,
  and $p_a$ is interpreted as the momentum flowing through the $a$-th $A$-cycle.

  Using the explicit expression (\ref{factorized}) for (\ref{A1}), 
  we find that the residue $A_N$ of the $n$-point function of the Liouville theory reduces to the following integral 
    \bea
    A_N
    &\propto&
           \prod_{a=1}^g \int_{-\infty}^{+\infty} d p_a \left| C(w, m, \xi, p)
           \exp \left( 2 \pi i \sum_{a, b} p_a p_b \tau_{ab} \right) \right|^2
           \nonumber \\
    & &    \prod_{i=1}^N \int d^2 z_i |\omega(z_i)|^{2+2b^2} \left| \exp \left( 4 \pi  b \sum_{a, i} p_a \int^{z_i} 
           \omega_a \right) \right|^2
           \nonumber \\
    & &    \left| \prod_{i<j} E(z_i, z_j)^{-2 b^2} \prod_{i, k} E(z_i, w_k)^{2bm_k} \prod_{I, i} E(\xi_I, z_i)^{-1-b^2} 
           \right|^2,
           \label{Amp_ctdt}    
    \eea
  where the factor $C(w, m, \xi, p)$ was defined as
    \bea
    C(w, m, \xi, p)
    &=&    \prod_I \omega (\xi_I)^{\frac{Q^2}{4}} \prod_{k=1}^n \omega (w_k)^{\frac{m_k^2}{g_s^2}}
           \prod_{k < \ell} E(w_k, w_\ell)^{- \frac{2 m_k m_\ell}{g_s^2}} \prod_{I, k} E(\xi_I, w_k)^{\frac{Q m_k}{g_s}}
           \label{C} \\
    & &    \times 
           \prod_{I<J} E(\xi_I, \xi_J)^{- \frac{Q^2}{2}} 
           \exp \left[ 2 \pi \sum_a p_a \left( Q \sum_I \int^{\xi_I} \omega_a
         - 2 \sum_k m_k \int^{w_k} \omega_a \right) \right].
           \nonumber
    \eea
  
  As in the torus case \cite{MY}, 
  it is not straightforward to factorize the integrals over the Riemann surface into
  holomorphic and anti-holomorphic integrals for generic $N$. 
  However this is easily performed in the large $N$ limit.
  Indeed, the last two-lines of (\ref{Amp_ctdt}) can be written as
  \be
  \int \prod_i d^2 z_i \,  \vert \mu {\rm e}^{\frac{b}{g_s}W}\vert ^2 
   \sim  \left\vert \int \prod_i dz_i \mu {\rm e}^{\frac{b}{g_s}W} \right\vert^2
  \label{straight}
  \ee
  where $\mu$ and $W$ are 
  \bea
  \mu &=& \left[\omega(z_i)^{1+b^2}  \prod_{i,I} E(z_i, \xi_I)^{-1-b^2}\right]
           \prod_{1 \le i < j \le N} E (z_{i}, z_{j})^{-2b^2}        
	   \prod_i E(z_i,z^*)^{2b\sum_k m_k/g_s}, \label{measure}\\
  W &=&  \sum_{i=1}^N 
           \left( \sum_{k=1}^n 2 m_k \log \frac{E (z_i,w_k)}{E(z_i,z^*)} 
           + 4 \pi  \sum_{a=1}^g p_{a} \int^{z_i} \omega_a  \right) \ ,  \label{c_action}
  \eea
    where we have chosen a base point $z^*$ in order to split the measure from the potential 
  and we have rescaled the parameters as $m_k \rightarrow m_k/g_s$ and $p_a \rightarrow p_a/g_s$.
  Notice that the term in the square brackets in (\ref{measure}) is independent on the zeroes of the conformal factor,
  ensuring therefore that the generalized matrix model correctly embodies the conformal symmetry of Liouville
  theory.
  
  The large $N$ limit amounts to take $g_s\to 0$ keeping $g_sN$, $b$, $m_k$ and $p_a$ finite.
  In this limit, the conditions for the criticality are given by
    \bea
    b \sum_{j \neq i}\frac{E'(z_{i}, z_{j})}{E(z_{i}, z_{j})} dz_i
    - \sum_{k=1}^n m_k \frac{E'(z_i, w_k)}{E(z_i,w_k)} dz_i
    - 2 \pi  \sum_{a=1}^g p_a \omega_a (z_i)
     = 0
           \label{ext_cond}
    \eea
  where $E'(z_1,z_2) \equiv \partial_{z_1} E(z_1,z_2)$.
  The conditions obtained from the $\bar{z}_i$-derivatives
  are just the complex conjugate of (\ref{ext_cond}).
  It is remarkable that the conditions for criticality are separated into
  holomorphic and anti-holomorphic equations,
  which implies that the integrals over the Riemann surface in (\ref{Amp_ctdt}) can be factorized
  into holomorphic and anti-holomorphic integrals in the large $N$ limit as stated in (\ref{straight}).
  We are therefore left with the following matrix-like integral
    \bea
 &&   Z^{{\cal C}_{g,n}}_N ({\tt w, m,p, v})
    \equiv   \nonumber\\
 &&          \int \prod_{i=1}^N dz_i \  \left[\omega(z_i)^{1+b^2}  \prod_{i,I} E(z_i, \xi_I)^{-1-b^2}\right]
           \prod_{1 \le i < j \le N} E (z_{i}, z_{j})^{-2b^2}        
	   \prod_i E(z_i,z^*)^{2b\sum_k m_k/g_s}
           \nonumber \\
    & &    \times \exp \left(  \frac{b}{g_s} \sum_{i=1}^N 
           \left( \sum_{k=1}^n 2 m_k \log \frac{E (z_i,w_k)}{E(z_i,z^*)} 
           + 4 \pi  \sum_{a=1}^g p_{a} \int^{z_i} \omega_a  \right) \right),
         \label{holo}
    \eea
  where ${\tt w} = \{ w_k\}$ , ${\tt m} =\{ m_k\}$, ${\tt p}=\{p_a\}$ and ${\tt v}=\{\nu_\alpha\}$ are the
  filling fractions $\nu_\alpha\equiv b g_s N_\alpha$ which specify the holomorphic integral above.
  The integrand in (\ref{holo}) is a proper one-from in each variable $z_i$ on the covering space of the Riemann surface
  due to momentum conservation (\ref{mom_cons}). The matrix model potential that we find is in the form anticipated by
  \cite{DV}.
 
 In order to count the number of moduli of our matrix model we should note that there are $n + 2 g - 3$ independent filling fractions:
  naively the number of critical points of the action is $2g - 2 + n + 1$.
  However, there are constraints coming from the fact that we are free to move the base point $z^*$
  and that we have specified the residue at the base point as above by using the momentum conservation.
  The latter is equivalent to the constraint on the sum of filling fractions $\sum_\alpha \nu_\alpha = bg_s N$.
  These constraints reduce the number of moduli by two, thus giving the correct counting.
  The paths of the integrals are defined such that 
  only the solution of (\ref{ext_cond}) labeled by the fixed filling fractions $\{ \nu_\alpha \}$
  contributes to the integrals.
  The measure factor in (\ref{holo}) can be regarded  
  as a generalization of the Vandermonde determinant.
  The differential $dz \partial_z W$ has simple poles 
  with residues $(\{2m_k\}, - 2 \sum_k m_k)$ at the points $(\{w_k\}, z^*)$.

  The integral in (\ref{Amp_ctdt}) is then obtained by integrating (\ref{holo}) 
  and its complex conjugate over the filling fractions.
  Thus, in the large $N$ limit, $A_N$ becomes
    \bea
    & &
           A_N
    =      \int^{\infty}_{-\infty} \prod_{a=1}^g dp_a \int \prod_{k=1}^{n+2g-3} d \nu_k 
           \Bigg|  
           \exp \left( 2 \pi i \sum_{a=1}^g \sum_{b=1}^g p_a \tau_{ab} p_{b} \right)
           C({\tt w, m, p}, \xi)
           Z^{{\cal C}_{g,n}}_N ({\tt w, m,p, v})
           \Bigg|^2  .
           \nonumber \\
    & &
           \label{FinalA}
    \eea
  At the level of the generalized matrix model, the filling fractions $\nu_\alpha$ are free parameters.
  Together with $p_a$ ($a = 1,\dots, g$) which are independent parameters in the potential,
  we have totally $n + 3 g -3$ independent moduli 
  which are identified with the internal momenta 
  in the Liouville conformal block and then with 
  the Coulomb moduli of the gauge theory.
  The explicit correspondence of these parameters with the internal momenta 
  is discussed in detail in section \ref{subsec:spectral_coulomb}.
  Under this identification, we see from (\ref{FinalA}) that  $ Z^{{\cal C}_{g,n}}_N ({\tt w, m,p, v})$
  is proportional to the conformal block of Liouville theory \cite{BG}.

\section{Degenerations}
\label{sec:degeneration}

To study the behavior of the generalized matrix model when approaching perturbative corners
in the space of gauge couplings, we have to study what happens when we degenerate ${\cal C}_{g,n}$.
The degeneration is usually described by using the plumbing fixture decomposition of the curve.
Let ${\cal U}_t$ be the annulus
$$
{\cal U}_t=\left\{
(z,w)\vert zw=t; |t|<|z|<1; |t|<|w|<1
\right\}
$$
which as $t\sim0$ describes the squeezed cylinder.
The curve undergoes the decomposition ${\cal C}_{g,n} = {\cal C}_{g-1,n,2}\cup {\cal U}_t$
when the degeneration is of pinching type and 
${\cal C}_{g,n} = {\cal C}_{g_1,n_1,1}\cup {\cal U}_t\cup {\cal C}_{g_2,n_2,1}$ with $g_1+g_2=g$
and $n_1+n_2=n$, with $1-2g_i-n_i<0$,
when the degeneration is dividing. The components ${\cal C}_{g,n,h}$ are here Riemann surfaces 
with genus $g$, $n$ punctures and $h$ non overlapping disks removed which will become the 
punctures in the degeneration limit.
The fact that the holomorphic integrals react correctly under the degeneration of the curve 
${\cal C}_{g,n}$
is a remnant of the analogous property of the conformal field theory
\cite{vafa-punct}
and is indeed a consequence of the construction we performed in the previous section.
We assume the shrinking cycle do not intersect the contour system along which (\ref{holo}) is evaluated.

Let us focus on the dividing case first. In this case the prime form $E(z,z')$ behaves as follows.
If both its arguments belong to a given same component, the prime form reduces to the prime form 
on that component,
while if its arguments belong to different components, then 
$E(z',z'')\sim E_1(z',P_1) E_2(P_2,z'') t^{-1/2}$, where $P_{1,2}$ correspond to the punctures created by the dividing.
To see what the prime form degeneration implies for the generalized matrix model measure and 
potential,
we have to split the integration contours in components according to the dividing 
decomposition.
This splits the $\{z_i\}$ in two sets according to 
which components of the contour they are integrated along, namely $N'$ of them on the first 
component and $N''$ on the second with $N=N'+N''$.
Correspondingly, also the puncture set will split in two subsets 
${\tt w}={\tt w'}\cup{\tt w"}$, one for each component.
By using the above degeneration formulas for the prime form and the fact that 
the holomorphic harmonic differentials $\omega_a$ reduce to the ones 
relative to the two splitting factors, we get that
\be
Z_N^{{\cal C}_{g,n}}({\tt w, m,p, v})
\sim
Z_{N'}^{{\cal C}_{g_1,n_1+1}}({\tt w'}\cup P_1, {\tt m'}\cup m^*_1,{\tt p', v'})
Z_{N''}^{{\cal C}_{g_2,n_2+1}}({\tt w''}\cup P_2, {\tt m''}\cup m^*_2,{\tt p'', v''})
\label{div_m}\ee
where 
$m^*_1=b g_s N'-\sum_{k'}m_{k'} + g_s Q(g_1-1)$
and 
$m^*_2=b g_s N''-\sum_{k''}m_{k''} + g_s Q(g_2-1)$ after using momentum conservation.
In the computation of (\ref{div_m}) one needs to count the two extra zeros for the reference holomorphic differential
to be placed at the location of the two resulting punctures.
The direct computation of the above mass formulas from the integral (\ref{holo})
indeed gives, for example, $m^*_1=-bg_s N''-g_2 g_s Q+\sum_{k''}m_{k''}$
where
the first term comes from the generalized Vandermonde,
the second from the measure term in the square bracket
and the third from the punctures.
The computation of $m^*_2$ is identical.
Notice that $m^*_1+m^*_2=-g_sQ $ corresponding to the fact that the two Liouville insertions 
generated at the 
punctures are conjugated and therefore the masses of the two flavors at the two punctures 
$P_i$ are equal.
The conformal modulus which gets traded to the mass is the total filling fraction
between the two integrals.
Formula (\ref{div_m}) is valid for each dividing degeneration such that
$g=g_1+g_2$, $n=n_1+n_2$ with $1-2g_i-n_i<0$.

In the pinching case, the prime form restricts to the one of the degenerate surface together
with the holomorphic differentials with zero $\alpha_g$-cycle. The left over
one scales as $2 \pi i \omega_g(z)\sim \partial_z \log\frac{E(z,P_1)}{E(z,P_2)}$
up to $O(t)$ terms.
In this limit, since $\omega$ in the conformal factor has been chosen to be regular, two of its zeroes will be at the two 
punctures generated at the pinching node and by direct computation one gets
\bea
Z_N^{{\cal C}_{g,n}}({\tt w, m,p, v})
\sim 
Z_N^{{\cal C}_{g-1,n+2}}({\tt w}\cup \{P_1, P_2\},{\tt m}\cup \{m^*_+ , m^*_-\} ,{\tt \hat p, v})
\label{pinch_m}\eea
where
$\hat p_a$ are the momenta in the $g-1$ left over handles and
\bea
m^*_\pm=-\frac{g_s Q}{2} \pm i p_g,
\label{pinchingmass}
\eea
with $p_g$ the momentum
in the squeezed one.
The two contributions to the above mass formulas arise respectively from the 
term in the square brackets in the measure 
and
the second term in the potential.
Once again, the two masses at the generated punctures are Weyl conjugated
$m^*_+ + m^*_-= - g_s Q$.
In the pinching case the conformal modulus which is traded for the mass is 
the momentum flowing in the squeezed handle.
The formulas above are general and valid at finite $N$.

In the following subsection we will discuss in detail the punctured torus case as an illustration.
One could also study for example the punctured genus two case. This case is special since all
degenerations reduce to punctured tori
\footnote{
The dividing degeneration 
${\cal C}_{2,n}\to {\cal C}_{1,n'+1}\cup {\cal C}_{1,n-n'+1}$
generates two punctured tori. Indeed, the genus two prime form in such a degeneration 
reduces to the relevant $\theta$-functions on the two tori since the period matrix 
at genus two becomes diagonal in the degeneration limit.
In the pinching case ${\cal C}_{2,n}\to {\cal C}_{1,n+2}$, the genus two $\theta$-function entering the explicit expression of the
prime from 
as $E(z,w)=\frac{\theta\left(\int_w^z\overrightarrow{\omega},\tau\right)}{\sqrt{\omega_\square(z)}\sqrt{\omega_\square(w)}}$
contracts to the torus $\theta$-function times a contribution from the off-diagonal term
of the period matrix which cancels in the degeneration limit the square-roots of the abelian differentials 
appearing in the denominator of the prime form.}. We discuss some aspects of the Seiberg-Witten geometry
on genus two curves in the Appendix.

\subsection{Degenerations of punctured tori}
\label{subsec:torus}
  In this subsection, we concentrate on the pinching degeneration of a torus 
  which leads to a sphere with two more punctures.
  Associated with the torus with $n$ punctures we can consider a class of quiver gauge theories \cite{Witten, Gaiotto}
  whose particular weak coupling descriptions include a gauge theory with circular quiver.
  Specifying a particular weak coupling description corresponds 
  to choosing a particular marking (a pants decomposition) of the Riemann surface.
  This gauge theory will reduce in the pinching of the torus to a linear quiver theory with $n-1$ $SU(2)$ gauge groups
  associated with a sphere with $n + 2$ punctures.
  In what follows, we verify that the generalized matrix model correctly reduces to 
  the Penner type matrix model on the sphere \cite{DV}.
  
  Since on the torus the canonical bundle is trivial, the choice of a base point is not needed.  
  The prime form is $E(z,w)=\frac{\theta_1(z-w|\tau)}{\theta'_1(0|\tau)}$ 
  and therefore the generalized matrix model (\ref{holo}) reduces to
    \bea
    Z^{{\cal C}_{1,n}}_N
     \sim     \int \prod_{i=1}^N d z_i \prod_{i<j} \theta_1(z_i-z_j)^{-2b^2} e^{\frac{b}{g_s} \sum_i W(z_i)},
           \label{torusZ}
    \eea
 up to $z_i$-independent factors.
 The potential is
    \bea
    W(z)
     =     \sum_{k=1}^n 2 m_k \log \theta_1 (z- w_k) + 4 \pi p z,
    \eea
  and     
    \bea
    \theta_1 (z)
     =     2 \sin (\pi z) \prod_{m=1}^\infty (1 - e^{2 \pi i z} q^m)(1 - e^{-2 \pi i z} q^m)(1 - q^m),
    \eea
  with $q = e^{2 \pi i \tau}$.
  The momentum conservation is given by
    \bea
    - \sum_{k=1}^n m_k + b g_s N
     =     0.
           \label{momentumtorusn}
    \eea
  Also, the identification of the moduli of the torus and 
  the gauge coupling constants of $SU(2)$ gauge groups $q_k$ is as follows \cite{AGT, MY}:
    \bea
    e^{2 \pi i (w_1 - w_2)}
     =     q_1, 
           ~~~
    e^{2 \pi i (w_2 - w_3)}
     =     q_2,
           ~\ldots, ~~
    e^{2 \pi i (w_{n-1} - w_n)}
     =     q_{n-1}, 
           ~~~
    e^{2 \pi i \tau}
     =     \prod_{k=1}^n q_k ,
    \eea
  which, by fixing $w_n = 0$, leads to
    \bea
    e^{2 \pi i w_{n-1}}
     =     q_{n-1}
     \equiv
           t_{n-1},
           ~~~
    e^{2 \pi i w_{n-2}}
     =     q_{n-2} q_{n-1}
     \equiv
           t_{n-2},
           ~~\ldots, ~
    e^{2 \pi i w_1}
     =     q_1 \dots q_{n-1}
     \equiv
           t_1.
           \label{torusgaugecoupling}
    \eea
  The mass parameters and one of the Coulomb moduli correspond to $m_k$ and $p$ respectively.
  
  Let us consider the pinching degeneration of the torus.
  We take $\Im \tau \rightarrow \infty$ which corresponds in the gauge theory to the decoupling limit
  of the $n$-th gauge group $q_n \rightarrow 0$.
  To consider the behavior of the generalized matrix model in this limit, 
  we first observe that the prime form reduces as
    \bea
    (dz)^{-1/2} (dw)^{-1/2} \frac{\theta_1(z-w)}{\theta_1'(0)}
    &\rightarrow&
           (dz)^{-1/2} (dw)^{-1/2} \frac{\sin \pi(z - w)}{\pi}
           \nonumber \\
    &=&    (d \xi)^{-1/2} (d \zeta)^{-1/2} (\xi - \zeta),
    \eea
  where in the last line we have changed coordinates to $\xi= e^{2 \pi i z}$ and $\zeta = e^{2 \pi i w}$.
  It is straightforward to see that the Vandermonde determinant of (\ref{torusZ}) reduces to 
  that of the usual $\beta$-deformed matrix model.
  The potential also reduces to
    \bea
    W(\xi)
     =     \sum_{k = 1}^{n} 2 m_k \log (\xi - t_k) + 2( - g_s Q/2 - i p) \log \xi,
           \label{torusactionpinching}
    \eea
  where we have used (\ref{torusgaugecoupling}) with $t_n = 1$. 
  Note that the first term corresponding to the momentum at $\xi = 0$ comes from the measure factor 
  $\omega(z)^{1+b^2}$ of the generalized matrix model (\ref{holo}).
  By the pinching, the punctures at $\xi = 0$ and $\infty$ are created.
  However, the latter disappeared from the potential, which thus
  reduces exactly to the Penner type matrix model \cite{DV, EM2}
  \footnote{The convention here is slightly different from the one in \cite{DV, EM, EM2}.
                           The momenta are related as $2 m = m_{{\rm DV}}$.
                           Our convention leads to the momentum conservation (\ref{momentum}).}
    \bea
    W(z)
     =     \sum_{k=0}^{n-1} 2 m_k \log (z - t_k) + 2 m_n \log (z -1),
           \label{actionspheren+2}
    \eea
  with the momentum conservation
    \bea
    - \sum_{k=0}^{n} m_k - m_\infty + b g_s N
     =     g_s Q,
           \label{momentum}
    \eea
  where $m_\infty$ is the momentum inserted at infinity.
  The relation between the parameters $t_k$ and the gauge couplings \cite{EM2} (See also \cite{Alba:2009ya})
  is the same as the one defined in (\ref{torusgaugecoupling}) with $t_0 = 0$.
  It follows from (\ref{torusactionpinching}) that $m_0 = - g_s Q/2 - i p$.
  
  Let us then analyze the momentum conservation under this degeneration.
  On one hand, in the original generalized matrix model, the conservation is described by (\ref{momentumtorusn}).
  On the other hand, in the Penner type one, the conservation is (\ref{momentum}).
  The momentum at infinity is then
    \bea
    m_{\infty}
     =   - \frac{g_s Q}{2} + i p.
    \eea
  These values of the momenta $m_{0}$ and $m_{\infty}$ are the ones which were already derived 
  in the generic analysis of the previous section (\ref{pinchingmass}).
  Note that there is a slight difference between the momenta $m_0$ and $m_\infty$, 
  which however disappears in the large $N$ limit.
  
  The original generalized matrix model has $n-1$ independent filling fractions $\nu_\alpha$. Recall that
  the overall $\sum_\alpha \nu_\alpha = b g_s N$ is constrained by the momentum conservation.
  Thus, by adding $p$, we have $n$ independent parameters which are identified 
  with the vevs of the vector multiplet scalars.
  The degeneration limit and the above argument mean that 
  $p$ in the potential is the vev of the $n$-th $SU(2)$ vector multiplet scalar
  and some combinations of the filling fractions are the vevs of the other $SU(2)$ scalars.
  We will give the precise identification in the large $N$ limit in the next section.

\section{Spectral curve of the generalized matrix model}
\label{sec:spectral}
  In this section, we derive the spectral curve of the generalized matrix model (\ref{holo}) in the large $N$ limit
  and show that it coincides with the Seiberg-Witten curve of the corresponding gauge theory.
 
  In the large $N$ limit, the evaluation of (\ref{holo}) 
  reduces to the calculation of the critical points.
  The condition for criticality is given by
    \bea
     d W(z_i) 
           - 2b g_s \sum_{j\neq i} d_{z_i} \log \left( \frac{E (z_{i}, z_{j})}{E (z_{i}, z_*)} \right) = 0, 
    \label{ex_con}
    \eea
  where the potential $W(z)$ is defined in (\ref{c_action})
  and we have used the momentum conservation (\ref{mom_cons}).
  Then, the prepotential in the large $N$ limit, defined as $\exp \left( {\cal F}/{g_s^2} \right) \equiv Z$,
  is given by
    \bea
    \frac{1}{g_s^2} {\cal F}
     =     \frac{b}{g_s} \sum_i W(z_i) 
         - 2b^2 \sum_{i<j} \log\left( \frac{E (z_{i}, z_{j})}{E (z_{i}, z^*)} \right),
             \label{pre}
    \eea
  where each ``eigenvalue'' $z_i$ satisfies (\ref{ex_con}).
 
  It is natural to assume that the eigenvalues 
  are distributed in line segments around the critical points of $W(z)$,
  similarly to the usual matrix model. Indeed the second term in (\ref{pre}) reduces locally to the standard
  Coulomb gas potential.
  We denote the line segments as $C_\alpha$ where $\alpha=1,\cdots, n + 2g - 2$.
  We assume that $C_\alpha$ do not include the base point $z^*$
  and the punctures $w_k$, at which the potential $W(z)$ diverges.
  We denote by $N_\alpha$ the number of eigenvalues on the line segment $C_\alpha$,
  where $N_\alpha$ satisfies $\sum_{\alpha=1}^{n+ 2g - 2} N_\alpha = N$. 

 Let us introduce the eigenvalue density current $\rho(z)$ 
  supported on $\{C_\alpha\}$ and normalized as
 $\oint_{C_\alpha}
  \rho (z) = b g_s N_\alpha \equiv \nu_\alpha$.
  Using the variables introduced above, the prepotential and the condition
  for criticality are written as
    \bea
    &&{\cal F} 
     =     \int_{\sum_\alpha C_\alpha}
           \rho (z) W(z)
           - \int_{\sum_\alpha C_\alpha} 
            \int_{\sum_\alpha C_\alpha} 
            \rho (z) \rho (z') 
           \log \frac{E(z, z')}{E(z,z^*)}
           \label{prepot}, \\
    &&  
       d W(z) - 2 \int_{\sum_\alpha C_\alpha} \rho (z') 
       d_z \log \left( \frac{E(z, z')}{E (z, z^*)} \right) =0,
           \label{cond_extrem}
    \eea
  respectively. Here, $z$ in (\ref{cond_extrem}) is on either of the 
  line segment $C_\alpha$ and the integral is defined as the principal integral.

  In order to solve the above condition (\ref{cond_extrem}) , we define the following one form,
  which is the generalization of the resolvent of the usual matrix model
    \bea
    R(z) 
    \equiv \int_{\sum_\alpha C_\alpha} 
    \, \rho (z') d_z 
           \log \left( \frac{E (z, z')}{E (z, z^*)} \right) .
          \label{resolvent}
    \eea
  This ``resolvent'' is defined at generic points $z$ on the Riemann surface
  contrary to the second term in (\ref{cond_extrem}).
  Note that the resolvent as well as $dW(z)$
  are single-valued one-forms on the Riemann surface.
  The resolvent has cuts at the line segments $C_\alpha$ and a simple pole at $z^*$.
  Also, the filling fractions are obtained by integrating the resolvent along the cuts as
    \bea
    \nu_\alpha 
     =     \frac{1}{2\pi i} \oint_{C_\alpha} R(z).
           \label{filling}
    \eea
  On the line segments $C_\alpha$, the resolvent behaves as
    \bea
    &&R(z + i \varepsilon e^{i\varphi(z)}) + R(z - i \varepsilon e^{i\varphi(z)}) 
     =     2 {\rm P} \int_{\sum_\alpha C_\alpha} 
     \rho(z') 
           d_z \log \left( \frac{E(z, z')}{E(z,z^*)} \right)
     =     d W (z),
           \label{prin} \\
    &&R(z + i \varepsilon e^{i\varphi(z)}) - R(z - i \varepsilon e^{i\varphi(z)}) 
     =     \oint_{z} 
     \rho(z') d_z \log \left( \frac{E(z, z')}{E(z,z^*)} \right)
     =     - 2 \pi i \rho (z) 
           \label{cut} 
    \eea
  where we take the real number $\varepsilon$ infinitely small 
  and we assume that a properly defined function $\varphi(z)$ exists such that 
  $z + i \varepsilon e^{i\varphi(z)}$ or $z - i \varepsilon e^{i\varphi(z)}$
  does not go across the cuts $C_\alpha$ when $z$ moves along $C_\alpha$. 
  The integral in (\ref{prin}) is principal integration,
  which is given as an average of integral 
  along the path above the singularity and that below the singularity.
  The resolvent should be determined such that (\ref{prin}) and (\ref{cut}) 
  are satisfied for $z \in C_\alpha$.
  A candidate of the solution for (\ref{prin}) is
    \bea
    R_0(z) 
     =     \frac{1}{2} d W(z).
    \eea
  However, it does not reproduce the correct structure of singularity expressed in (\ref{cut}).
  We need singular contributions:
    \bea
    R(z) 
     =     \frac{1}{2} d W(z) + R(z)_{\rm sing}, 
           \label{R_def}
    \eea
  where (\ref{prin}) and (\ref{cut}) impose
    \bea
    R(z + i \varepsilon e^{i\varphi(z)})_{\rm sing} + R(z - i \varepsilon e^{i\varphi(z)})_{\rm sing}
    &=&     0.
            \label{sing_prin}
    \\
    R(z + i \varepsilon e^{i\varphi(z)})_{\rm sing} - R(z - i \varepsilon e^{i\varphi(z)})_{\rm sing}
    &=&     - 2 \pi i \rho (z) .
            \label{sing_cut} 
    \eea
  The above discussion is valid for a generic potential $W(z)$.
  In the following, we use its explicit form (\ref{c_action}) to determine the resolvent $ R(z) $.
  Then, we find that
    \bea
    R_{\rm sing} (z)
    &=&
        \int_{\sum_\alpha C_\alpha} \rho (z') 
           d_z \log \left( \frac{E (z, z')}{E (z, z^*)} \right) 
           - \sum_{k=1}^n m_k d_z \log \left( \frac{E (z, w_k)}{E(z, z^*)} \right) 
           - 2 \pi \sum_{a=1}^g p_a \omega_a (z)
           \cr
    &=&  \int_{\sum_\alpha C_\alpha} 
      \rho (z') d_z \log E (z, z') 
           - \sum_{k=1}^n m_k d_z \log E (z, w_k) 
           - 2 \pi \sum_{a=1}^g p_a \omega_a (z)
    \label{noname}
    \eea  
  does not depend on the base point $z^*$,
  where we used the momentum conservation (\ref{mom_cons})
  and ignored the subleading term in the large $N$ expansion.
  We see that $R(z)_{\rm sing}$ has cuts in the regions $C_\alpha$
  and simple poles with residues $m_k$ at $z=w_k$. Moreover, it is independent on
  the base point $z^*$ as expected.

  From (\ref{sing_prin}), we see that the sign of $R(z)_{\rm sing}$ changes 
  across the cuts. Therefore its square only displays singularities at the
  punctures $z=w_k$. From (\ref{noname}) we see that these are at most quadratic poles
  with coefficients $m_k{}^2$.
  The spectral curve of the generalized matrix model thus reads
    \bea
    R_{\rm sing} (z){}^2 
     =     \sum_{k=1}^n m_k{}^2 \eta(z,w_k) + \zeta(z)
           \label{scurve}
    \eea
  where $\eta(z,w_k)$ are quadratic Strebel differentials, with double pole at $w_k$,
  and $\zeta (z)$ is a quadratic differential which has at most simple poles at $w_k$.
  $\zeta(z)$ is determined in terms of $n+3g-3$ parameters; in particular it depends on the $n+2g-3$ 
  independent filling fractions $\nu_\alpha$ and the $g$ internal momenta $p_a$.
  This form of the curve is the same as that of the Seiberg-Witten curve of quiver gauge theory as
  a cover of the base Riemann surface \cite{Witten,Gaiotto}. 
  As discussed in \cite{DW}, the physical information is included in the Prym variety of the Seiberg-Witten curve,
  rather than the Jacobian variety.
  This reduces the number of independent periods of the Seiberg-Witten differential to $n + 3g - 3$,
  which agrees with the number of the filling fractions and $p_a$.
 
  In order for the above argument to be a check of AGT correspondence at all genera,
  we need to show that the spectral curve is indeed the same as the one proposed, 
  in the Virasoro conformal block side, to be the Seiberg-Witten curve: 
  the insertion of the energy-momentum tensor in the conformal block 
    \bea
    \left< T(z) \prod_{k=1}^n V_{\alpha_k} (w_k) \right>_{{\mathcal C}_{g}}
     \rightarrow 
         - \frac{x(z)^2}{g_s^2} \left< \prod_{k=1}^n V_{\alpha_k} (w_k) \right>_{{\mathcal C}_{g}}
           \label{virasoroT}
    \eea
  in the semi-classical limit, which corresponds to the large $N$ limit in the generalized matrix model.
  It was claimed that $x dz$ coincides with the Seiberg-Witten differential.
  This was checked in the cases on a sphere and a torus already in \cite{AGT}.
  A useful way to capture the energy-momentum tensor insertion is 
  to consider the insertion of the degenerate field 
  $\Phi_{1,2}(z)\equiv e^{- \frac{\tilde{\phi}(z)}{b}}$ in the conformal block.
  By this insertion, the conformal block satisfies 
  the BPZ equation and can be expanded in $g_s$ as \cite{AGGTV}
    \bea
    \left< \Phi_{1,2} (z) \prod_{k=1}^n V_{\alpha_k} (w_k) \right>_{{\mathcal C}_{g}}
     \sim
           \exp \left( \frac{\CF_0}{g_s^2} + \frac{1}{b g_s} \int^z x(z') dz' + \ldots \right)
           \label{VirasoroT2}
    \eea
  Thus, the counterpart of the Seiberg-Witten differential in the Virasoro side can be obtained by calculating
    \bea
    x(z) = b g_s \frac{\partial}{\partial z} \log                                                  
    \frac{ \left< \Phi_{1,2} (z) \prod_{k=1}^n V_{\alpha_k} (w_k) \right>_{{\mathcal C}_{g}}}%
    { \left<\prod_{k=1}^n V_{\alpha_k} (w_k) \right>_{{\mathcal C}_{g}}}
    \label{SW_mat}
    \eea 
  in the semi-classical limit. 
  
  In the following, we rewrite (\ref{SW_mat}) in terms of the generalized matrix model by
  using the discussion in section \ref{sec:Liouville}.
  The corresponding calculation in the case of the sphere has been done in \cite{MMMsurface}.
  We use the holomorphic half of the integrand in (\ref{FinalA}) 
  to evaluate the conformal block in (\ref{SW_mat}).
  The conformal block with the insertion is also obtained just by changing $n$ to $n+1$ and by
  regarding that $m_{n+1}= \frac{g_s}{2b}$ and $w_{n+1}=z$.
  The momentum conservation (\ref{mom_cons}) is slightly modified to
    \bea
    \sum_{k=1}^n m_k + \frac{g_s}{2b} + g_s Q (1-g) 
     =     b g_s N.
           \label{momentumCgn+1}
    \eea
  By collecting the factor dependent on $z$, we obtain 
    \bea
   x(z) 
    &=& bg_s \frac{\partial}{\partial z}
           \log \left( \omega(z)^{\frac{1}{4b^2}} \prod_I E(z, \xi_I)^\frac{Q}{2b}
           E(z, z^*)^{N - \frac{\sum_k m_k}{bg_s}}
         e^{- \frac{W(z)}{2 b g_s}} \left< \prod_i \frac{E(z, z_i)}{E(z, z^*)} \right> \right) \cr
    &=&   - \frac{1}{2} \frac{\partial W(z)}{\partial z}
           + bg_s \frac{\partial}{\partial z} \log \left< \prod_i \frac{E(z, z_i)}{E(z, z^*)} \right> + {\cal O} (N^{-1}),
          \label{ratio}
    \eea
  where we have used the deformed momentum conservation (\ref{momentumCgn+1}) in the second equality.
  The expectation value is defined by 
    \bea
    \left< \ldots \right> 
      = \frac{\int \prod d z_i \mu e^{\frac{b}{g_s}\sum_i W(z_i)} \ldots}{\int \prod dz_i \mu e^{\frac{b}{g_s}\sum_i W(z_i)}}.
      \label{def_exp}
    \eea
  A subtlety is that the numerator in (\ref{def_exp}) is defined 
  with the deformed momentum conservation (\ref{momentumCgn+1})
  while the denominator with the original momentum conservation.
  However, note that the effect of the insertion of the degenerate field is a subleading contribution
  in the large $N$ limit, as in (\ref{VirasoroT2}).
  Note also that the deformation of the momentum conservation gives rise to that of the prepotential $\CF_0$,
  which does not depend on $z$ and disappears by the partial derivative in terms of $z$.
  Thus, we can use the same external momenta $m_k$ and the same $N$ as those of the denominator of (\ref{def_exp})
  to evaluate the numerator in the expectation value. 
  Also, some of the internal momenta in the numerator are shifted by $\pm b/2$ 
  depending on where we insert the degenerate operator as discussed in \cite{AGGTV},
  but again, the effect of this shift does not produce the factor dependent on $z$ in the large $N$ limit.
  Thus, the expectation value in (\ref{ratio}) can be evaluated by substituting the solution
  of the condition of criticality (\ref{ex_con}) or (\ref{cond_extrem}), which leads to
    \bea 
    bg_s \frac{\partial}{\partial z} \log \left< \prod_i \frac{E(z, z_i)}{E(z, z^*)} \right>
       =  bg_s \left< \sum_i  \frac{\partial}{\partial z} \log \frac{E(z, z_i)}{E(z, z^*)} \right>
          + {\cal O} (N^{-1}).
    \eea
  From the discussion above, we finally obtain
    \bea
    x(z) dz 
         \sim - \frac{dW(z)}{2} + R(z) 
         =  R_{{\rm sing}}(z),
    \eea
  up to $\CO(N^{-1})$ terms.
  This shows that the resolvent $R_{{\rm sing}}$ is indeed $x(z) dz$ in (\ref{virasoroT}),
  which was claimed to agree with the Seiberg-Witten differential.

\subsection{Dependence on internal momenta}
\label{subsec:spectral_coulomb}
  In this subsection, we show that the spectral curve of the generalized matrix model depends
  on the Coulomb moduli parameters in the same way as the Seiberg-Witten curve,
  which completes the check of AGT correspondence at all genera.
  In order to do that, we first have to specify 
  the marking of the Riemann surface ${\mathcal C}_{g,n}$,
  which is done by choosing $n+3g-3$ physically independent cycles.
  This marking determines the conformal block labeled by a trivalent graph on one hand, 
  and also the corresponding weak coupling description of the gauge theory on the other hand.
  On the gauge theory side, 
  each vacuum expectation value of each vector multiplet scalar is obtained by
  an integral along each cycle of the marking $\gamma_r$:
    \bea
    a_r
     =     \frac{1}{2 \pi i} \oint_{\gamma_r} \lambda_{{\rm SW}},
    \label{a_r}
    \eea
  where $\lambda_{{\rm SW}}$ is the Seiberg-Witten differential.
  We would like to show that the spectral curve also satisfies the corresponding relation.

  Before going to that issue, 
  we show in the following discussion that each pair of pants has one cut under the marking. 
  In general, three-punctured sphere always has one cut \cite{mm1} because
  the classical potential of the corresponding Penner type matrix model
  with generic mass parameters has one extremum.
  We can also show this by considering the dividing limit of four-punctured sphere:
  under the decomposition, it is known that 
  one cut exists in the pants including the punctures originally at $0$ and $q$,
  while the other cut exist in the pants including the punctures at $1$ and $\infty$.
  Furthermore, unless the mass parameters vanish, there are no punctures belonging to the cut.
  This indicates that three-punctured sphere with generic mass parameters has one cut apart from the three punctures.
  Since we have shown that our generalized matrix model behaves correctly in the degeneration limit,
  the cuts of its resolvent should be placed in such a way that each pair of pants has one cut.

  Once the marking is specified, an explicit correspondence should be determined 
  between the $n+3g-3$ internal momenta of the Virasoro conformal block
  and the $n+2g-3$ independent filling fractions together with the $n$ parameters $p_a$ in the potential.
  In order to see this explicitly,
  we can use the results on the degeneration obtained in section \ref{sec:degeneration}.
  Note that among the intermediate states in the conformal block,
  only the primary state remains and appears as the external state in the degeneration limit.
  Thus, the external momentum appeared by the degeneration which shrinks a particular cycle 
  can be seen as the internal momentum of the corresponding place before the degeneration.
  In order to determine all the internal momenta, 
  we pinch all the $A$-cycles of the Riemann surface ${\cal C}_{g,n}$ 
  until it becomes ${\cal C}_{0,n+2g}$
  and then divide ${\cal C}_{0,n+2g}$ until they split into three-punctured spheres
  in a way that it reproduces the specified pants decomposition.
  Whichever marking we take, we can read off all the internal momenta 
  by considering the corresponding degeneration limit.
 
  The $A$-cycles of $\mathcal{C}_{g,n}$ can be chosen to coincide with some of $\gamma_r$'s.
  Thus, the set $\{ \gamma_r \}$ can be divided as $\{ \gamma_r \} = \{ A_a \} \cup \{ \gamma_\alpha \}$
  where $a = 1, \ldots, g$ and $\alpha = 1, \ldots, n + 2g -3$, 
  such that $A$-cycles of $\mathcal{C}_{g,n}$ are $A_a$ and $\gamma_\alpha$ are the remaining cycles.
  By shrinking $A_a$-cycles, we obtain (\ref{pinchingmass}).
  Especially, in the large $N$ limit, the internal momenta (multiplied by $g_s$) 
  corresponding to the $A_a$ cycles of the Riemann surface ${\cal C}_{g}$ are given by 
  \bea
  a_a = ip_a. \quad (a=1,\cdots, g)
  \label{internal1}
  \eea 
  After pinching all the $A_a$ cycles,
  the remaining internal momenta can be obtained as written just below (\ref{div_m})
  by further considering the dividing limit which shrinks $\gamma_{\alpha}$ cycles.
  Since each pair of pants has one cut, on which $N_{\beta}$ eigenvalues $\{ z_i \}$ exist,
  the number of eigenvalues $N'$ or $N''$ on each side of the dividing component of the Riemann surface 
  can be written in terms of sum of the filling fractions. 
  In the large $N$ limit, the internal momenta (multiplied by $g_s$) 
  flowing through $\gamma_{\alpha}$ cycles are given by 
  \bea
  a_{\alpha} = \sum_{\beta} \nu_{\beta} - \sum_{k}m_{k} + \sum_a (\pm ip_a), \quad (\alpha=1,\cdots, n+2g-3)
  \label{internal2}
  \eea
  where the range of the sum depends which $\gamma_\alpha$ cycle we shrink
  under the marking of the Riemann surface.
  The third term is the contribution from new punctures which appear by pinching $A_a$ cycles
  in the previous step.
  The sign $\pm$ is determined for each puncture according to the direction of the $A_a$ cycle.
  Although the internal momenta (\ref{internal1}) and (\ref{internal2}) 
  are obtained in the degeneration limit, 
  we assume that these relations hold for arbitrary 
  moduli parameters of the Riemann surface ${\cal C}_{g,n}$
  because the filling fraction $\nu_{\alpha}$ and the parameter $p_a$ 
  are independent of the moduli parameters.
  It is remarkable that the form of our generalized matrix model in the large $N$ limit
  is universal for any choice of markings of the Riemann surface.
  However, the identification of the parameters 
  with internal momenta depend on such choice,
  which reflects in the difference of the conformal blocks labeled by different trivalent graphs.
  
  So far, we have discussed the relation between 
  the internal momenta of the Virasoro conformal block and the independent filling fractions and parameters $p_a$.
  In the following, we check that the corresponding relation as (\ref{a_r})
  is reproduced from the spectral curve under these identification of the parameters.
  By using (\ref{noname}), we can explicitly calculate the $A_a$-cycle integrals as 
  \bea
  \frac{1}{2 \pi i} \int_{A_a} R_{{\rm sing}} = ip_a   \quad (a=1,\cdots, g)
  \label{intA}
  \eea 
  Note that the convention here is $\oint_{A_a} \omega_b = \delta_b^a$.
  Taking into account the identification of the parameters (\ref{internal1}), 
  we see that the corresponding relation as (\ref{a_r}) for $A_a$-cycle is confirmed.
  
  For the $\gamma_\alpha$ cycles, the check is not direct,
  because $\gamma_\alpha$ are not the contours $C_\alpha$ along the cuts of the spectral curve of the matrix model.
  Recall that there are indeed $n + 2g -2$ cuts in the spectral curve, 
  but the integral of the resolvent along one of them is not independent due to the momentum conservation.
  We will call the remaining $n + 2g -3$ cuts the independent ones. 
  It follows that the cycles around the independent cuts, 
  which we have denoted by $C_{\alpha}$ ($\alpha=1,\cdots n+2g-3$),
  can be expressed as sum of the three cycles which go around each leg of the pants.
  That is, $C_{\alpha}$ can be written as a linear combination of $\gamma_r$ ($r =1,\cdots n+3g-3$)
  and $D_k$ ($k=1,\cdots n$), where $D_k$ are the contours around the simple poles of $R_{{\rm sing}}$.
  Thus, $\gamma_\alpha$ ($\alpha=1,\cdots n+2g-3$) can be written as 
  $\gamma_\alpha = n_{\alpha}{}^a A_a + c_{\alpha}{}^{\beta} C_{\beta} + s_{\alpha}{}^k D_k$,
  where the coefficients $n_{\alpha}{}^a$, $c_{\alpha}{}^{\beta}$ and $s_{\alpha}{}^k$ are integers.
  (Note that in addition to them there could exist a cycle around the base point.
  However, when we consider $R_{{\rm sing}}$, this dependence disappears in the large $N$ limit, 
  as stated in (\ref{noname}).)
  In terms of these, we calculate the integral over $\gamma_{\alpha}$ cycles as 
    \bea
    \frac{1}{2 \pi i} \oint_{\gamma_\alpha} R_{{\rm sing}}(z)
    &=&     \frac{1}{2 \pi i} \oint_{\gamma_\alpha} \left[ R(z) - \sum_{k=1}^n m_k \omega_{w_k, z^*}(z)
         - 2 \pi \sum_{a=1}^g p_a \omega_a (z) \right]
           \nonumber \\
    &=&    \sum_\beta  c_{\alpha}{}^{\beta} \nu_\beta - \sum_k s_{\alpha}{}^k m_k + i \sum_a n_{\alpha}{}^a p_a ,
           \label{coulombfilling}
    \eea
  where $\omega_{x, y}(z) = d_z \log \frac{E(z, x)}{E(z, y)}$ is the Abelian differential of the third kind.
  
    \begin{figure}[t]
    \begin{center}
    \includegraphics[width=6cm]{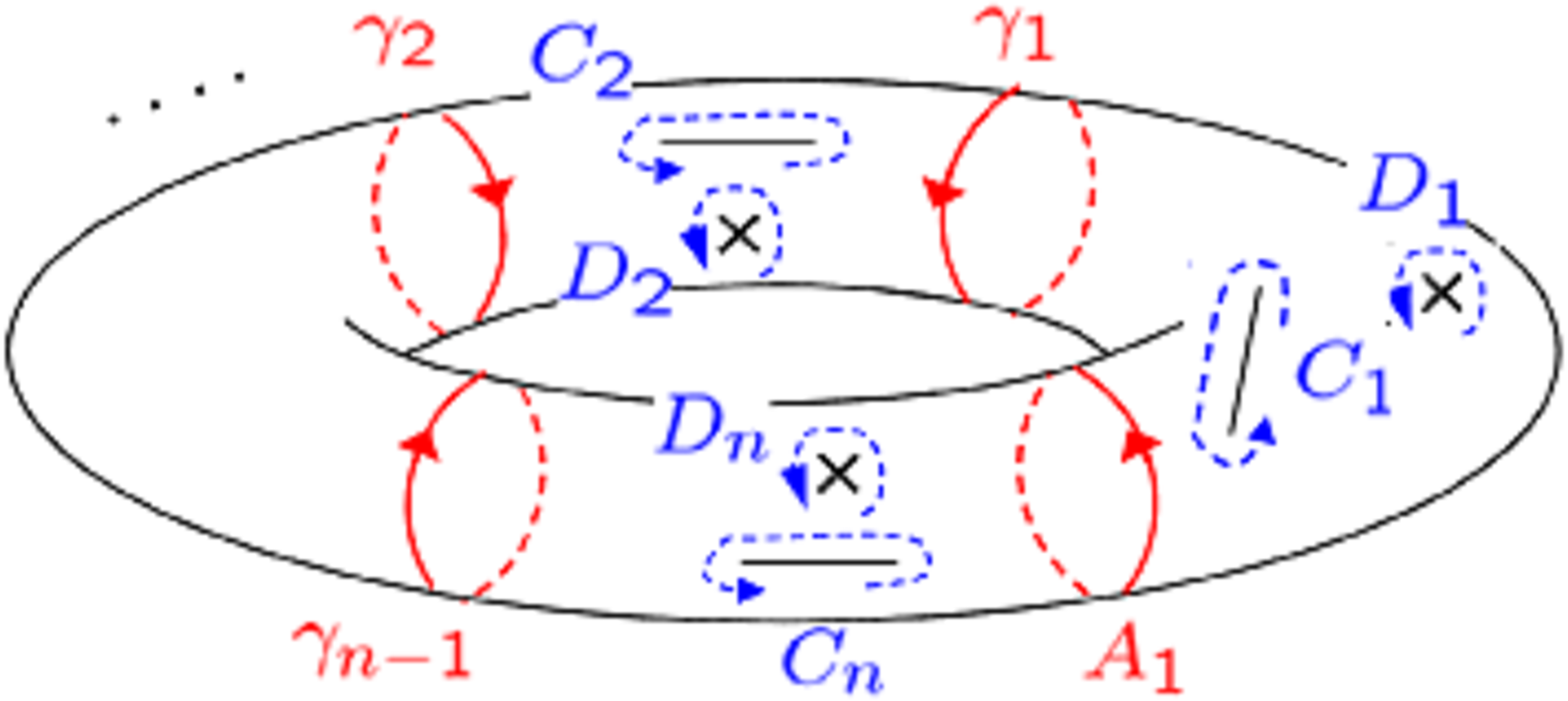}
    \caption{{\small A marking of $\mathcal{C}_{1,n}$ corresponding to the elliptic quiver 
                     and the $A_{1}$- and $\gamma_\alpha$-cycles.
                     E.g., $\gamma_1 = A_{1} + C_1 + D_1 $ and $\gamma_2 = A_{1} + C_1 + C_2 + D_1 + D_2$.}}
    \label{fig:C1n}
    \end{center}
    \end{figure}

  To determine the coefficients, we briefly show the example below.
  In the case of the torus $\mathcal{C}_{1,n}$, there is only one $A$-cycle and the number of independent cuts is $n-1$.
  For $n=1$, there is no independent cut and the parameter $p$ is identified 
  with the vacuum expectation value of the vector multiplet scalar of $\CN=2^*$ gauge theory as analyzed in \cite{MY}.
  For $n>1$, we consider the marking corresponding to $SU(2)^n$ elliptic quiver gauge theory
  and choose the $A_{1}$-cycle and $\gamma_\alpha$-cycles as depicted in Fig \ref{fig:C1n}.
  As stated above, each pair of pants has one cut.
  (This can also be explicitly checked by solving the equations of motion in the weak coupling limit.)
  Therefore, it is easy to obtain the coefficients above $n_\alpha^1 = 1$ and
    \begin{eqnarray}
    c_\alpha^\beta
     =    \left\{ \begin{array}{ll}
          1 & (\beta = 1, \ldots, \alpha) \\
          0 & (\beta = \alpha + 1, \ldots, n-1) \\
          \end{array} \right. ~~~
    s_\alpha^k
     =    \left\{ \begin{array}{ll}
          1 & (k = 1, \ldots, \alpha) \\
          0 & (k = \alpha + 1, \ldots, n-1) \\
          \end{array} \right.
    \end{eqnarray}
  This indicates
    \bea
    \frac{1}{2 \pi i} \oint_{\gamma_\alpha} R_{{\rm sing}}(z)
     =   \sum_{\beta = 1}^\alpha \nu_\beta - \sum_{k=1}^\alpha m_k + i p
    \eea
  in addition to the one obtained by the $A$-cycle integral (\ref{intA}).
  Comparing it with the result obtained from (\ref{internal2}), 
  we see that it exactly reproduces the internal momenta flowing through $\gamma_\alpha$ cycle.
  Similarly, it is straightforward to check for generic case 
  that (\ref{coulombfilling}) coincides with that obtained from (\ref{internal2}).
  Thus, we have confirmed that the corresponding relations as (\ref{a_r}) for $\gamma_{\alpha}$ cycles
  are reproduced.

\section{Conclusions}
\label{sec:conclusion}

In this paper, we have shown that 
the perturbative analysis of Liouville correlation functions 
displays in the large $N$ limit holomorphic factorization of the surface integrals
and leads to generalized matrix models, defined on the cover of ${\cal C}_{g,n}$, which describe the relevant 
Virasoro conformal blocks. We provided an all genera check of the AGT correspondence by 
obtaining the Seiberg-Witten data from the saddle point analysis of these generalized matrix models.

We underline that the models presented in this paper could be useful for the
exploration of the full set of gauge theories with generalized quiver structure of \cite{Gaiotto}. 
Indeed, so far most of the analysis of the AGT correspondence has been focused on the linear and elliptic quiver cases, 
mainly due to the lack of calculational tools for higher genera.
However, to fully exploit the generalized matrix model approach one should be able to extend its analysis to finite $N$.
This would amount to provide a full derivation at finite $N$ of the holomorphic factorization, which
in turn would give a precise prescription for the contour integrals possibly extending the recipe
of \cite{mm2,CDV} to higher genera.

Notice anyway that our large $N$ analysis depends only on the homotopy class of the contours
via the filling fractions as explained in Section 4.1.
In this sense our results are universal with respect to a particular choice of contour's representatives 
in the matrix integral.

Moreover, for gauge groups of higher rank, which according to \cite{DV,IMO} should correspond to multi-matrix models,
this approach could shed light on the description of  
strongly coupled sectors naturally appearing in the general framework and not admitting a known lagrangian description.

Another very interesting issue to explore is the relation of the generalized matrix models with the quantization of 
integrable systems \cite{integrable, MT}. (See also \cite{Eynard}.) In particular this could provide an alternative 
derivation of the quantum Hamiltonians for Hitchin integrable systems and generalize it to higher genera.

\section*{Acknowledgements}
A.T. thanks R.~Santachiara for useful discussions. 
F.Y. thanks T.~Eguchi and N.~Nekrasov for useful discussions.
G.B. is partially supported by the INFN project TV12. 
The research of K.M. is supported in part by JSPS Bilateral Joint Projects (JSPS-RFBR collaboration).
A.T. is partially supported by PRIN ``Geometria delle variet\'a algebriche e loro spazi di moduli'' and the INFN project PI14
``Nonperturbative dynamics of gauge theories''.

\appendix

\section*{Appendix}

\section{Addendum: playing with genus $2$ curves}

A hyperelliptic curve ${\cal C}_g$ of genus $g$ is given by the equation
$$
y^2=P_{2g+2}(x)
$$
where $P_{2g+2}$ is a polynomial of degree $2g+2$ and is realized 
\footnote{
Under conformal inversion $x=1/x'$ on the Riemann sphere, the stability of the description 
is guaranteed by the transformation $y=y' {x'}^{-(g+1)}=y'\left(\frac{\partial x}{\partial x'}\right)^{\frac{g+1}{2}}$.}
in the total space of $T\mathbb{P}^{\frac{g+1}{2}}$.

As it is well known, all genus 2 curves are hyperelliptic.
These are realized in general by a sextic polynomial equation
\be
y^2=\prod_{i=1}^6(x-a_i)
\la{sextic}\ee
which we denote by ${\cal C}_2$.

The complex structure moduli $\overline{\cal M}_{2,0}$ of genus two curves is then obtained 
by considering the complex parameters $\{a_i\}$
modulo the action of the permutation group $S_6$ and the ${\bf P}SL(2,{\bf C})$.

A basis of abelian differentials is given by $\omega_a=\frac{x^{a-1}dx}{y}$, $a=1,2$, while 
a basis of quadratic differentials is given by $\phi_\alpha=\frac{x^{\alpha-1}dx^2}{y^2}$, $\alpha=1,2,3$ \cite{FK}.

The Seiberg-Witten (SW) geometry of the $SU(2)$ theory at genus 2
is specified by a double cover of ${\cal C}_2$ in $T^*{\cal C}_2$. 
As such, this is specified by a general quadratic differential on ${\cal C}_2$ in the form
\be
w^2=\Phi_2
\la{sw}\ee
where $\Phi_2=\sum_\alpha K_\alpha\phi_\alpha$ can be expanded in the Coulomb moduli $K_\alpha$
of the theory.

The perturbative expansions of the theory are available in the vicinity of the degeneration 
locus of the moduli space, namely around 
\be
\partial\bar{\cal M}_{2,0}= \bar{\cal M}_{1,1}\times \bar{\cal M}_{1,1} \cup \bar{\cal M}_{1,2}
\la{deg}\ee
The second factor in (\ref{deg}) is still generically not lagrangian and has to be degenerated 
as $\partial\bar{\cal M}_{1,2}= \bar{\cal M}_{1,1}\times \bar{\cal M}_{0,3} \cup \bar{\cal M}_{0,4}$
to reach corners around which known lagrangian descriptions are available.
The first factor in (\ref{deg}), being given by two copies of the ${\cal N}=2^*$ $SU(2)$ theory, is already lagrangian.
The first degeneration is dividing and the second one is pinching.

Let us discuss the dividing case in detail.
This is reached by taking the limit in which three branch points in (\ref{sextic}) collide.
To be concrete, let's fix the position of two of them at $0$ and $\infty$,
write our curve as
\be
y^2=x(x-a_1\epsilon)(x-a_2\epsilon)(x-a_3)(x-a_4)
\la{normalized}\ee
and take the limit as $\epsilon\to0$.
The curve (\ref{normalized}) becomes in the $x$ coordinate
$$
y^2=x^3(x-a_3)(x-a_4)
$$
which, redefining $y=x\tilde y$, reads 
\be
\tilde y^2=x(x-a_3)(x-a_4)
\la{elliptic}\ee
that is the torus with a puncture at $x=0$.

Let us check now the SW geometry in the degeneration limit.
The issue to discuss is just the scaling of the Coulomb parameters in this limit.
Notice that the degeneration is obtained by contracting to zero two nearby branching points, therefore
we are saturating a complex structure modulus corresponding to a Beltrami differential $\mu_\epsilon$
with support around the origin of size $\sim \epsilon$. This is dual to a holomorphic quadratic differential
which, not to have a vanishing overlap integral with $\mu_\epsilon$ should not be zero at $x=0$.
This is uniquely determined to be $\frac{dx^2}{y^2}$. Therefore, along the limit with $\epsilon\to 0$, the corresponding 
parameter in the SW curve has to scale away.

As a consequence of the above reasoning, exposing the $\epsilon$-parameter, the SW curve is parametrized as
\be
w^2=\frac{u'\epsilon+m^2 x + ux^2}{y^2}(dx)^2
\la{SW}\ee
The degeneration of ${\cal C}_2$ is easily kept into account in the SW geometry which becomes
\be
w^2=\frac{m^2 x^{-1} + u}{\tilde y^2}(dx)^2
\la{inter}\ee
The standard parameterization of the punctured torus is in the coordinates where the puncture sits at $\infty$.
Therefore, we rewrite the elliptic curve (\ref{elliptic}) after the inversion $x=\frac{1}{x'}$
to pull the puncture at $x'=\infty$ and redefine accordingly $\tilde y= \frac{1}{(x')^2}{\tilde y}'$.
After this, the SW curve reads 
$$
w^2=\frac{m^2 x' + u}{({\tilde y}')^2}(dx')^2
$$
which we can put in the representation with respect to the periodic coordinate via the Weierstrass 
parameterization\footnote{The constant $c$ needs to bring (\ref{elliptic}) to the standard Weierstrass form where
the quadratic term vanishes and can be computed explicitly.}
$x'={\cal P}(z)+c$ and ${\tilde y}'=\frac{d}{dz}{\cal P}(z)$
so that we stay with
\be
w^2=\left[m^2{\cal P}(z) + (u+cm^2)\right](dz)^2
\la{n2star}\ee
which is the SW curve for a copy of the ${\cal N}=2^*$ theory.

The other copy corresponds to the other half in which the original genus 2 surface was split.
Let's see how to get this second copy. 
In order to do it we have to consider the curve in the coordinate appropriate for the 
other half, namely we have to change (\ref{normalized}) to $x=\epsilon/\hat x$ and correspondingly
$y=\hat y\frac{\epsilon^{3/2}}{(\hat x)^3}$ after which we get
\be
{\hat y}^2=\hat x (1-a_1\hat x)(1-a_2\hat x)(\epsilon-a_3\hat x)(\epsilon-a_4\hat x)
\la{hat} \, . 
\ee
(\ref{hat}) becomes after the degeneration limit the curve
$$
{\hat y}^2= {\hat x}^3(1-a_1\hat x)(1-a_2\hat x)(a_3a_4)
$$
which we bring to the form of a punctured torus by redefining $\hat y=\tilde{\hat y} \hat x$
and get
$$
{\tilde{\hat y}}^2= \hat x(1-a_1\hat x)(1-a_2\hat x)(a_3a_4) \ \ .
$$

Let's follow now what happens to the SW curve (\ref{SW}) in the $\hat x$ patch.
This becomes
\be
w^2=\frac{u'+m^2{\hat x}^{-1} + u\epsilon{\hat x}^{-2}}{{\hat y}^2}(d{\hat x})^2
\la{hatSW}\ee
which in the limit $\epsilon\to 0$ has the same form of (\ref{inter}), but referring to the second punctured torus  
with an independent Coulomb parameter $u'$.
So, following the same procedure leading to (\ref{n2star}), we get the second copy of ${\cal N}=2^*$
with an independent gauge coupling and Coulomb parameter but the same mass as the first.

A pinching of the genus $2$ curves (\ref{normalized}) can be obtained by letting $\epsilon\to\frac{a_3}{a_2}$ for example.
In such a case the curve gets to 
\be
y^2=x(x-a_1a_3/{a_2})(x-a_3)^2(x-a_4)
\label{pinched}\ee
which is, after renaming $y=\tilde y (x-a_3)$, the twice punctured torus
\be
\tilde y^2=x(x-a_1a_3/{a_2})(x-a_4).
\label{puncto}\ee

Correspondingly, the holomorphic quadratic differential entering the Seiberg-Witten curve (\ref{SW})
becomes
\be
\Phi_2\to \frac{u'a_3/{a_2}+m^2 x + ux^2}{(x-a_3)^2{\tilde y}^2}(dx)^2
\la{punctosw}\ee
which explicitly displays quadratic poles at the two images of $x=a_3$ with equal coefficients.
This coefficient is actually fixed by the Coulomb parameter corresponding to the ungauged 
group $SU(2)$ at the end of the shrinking of the handle.

\end{document}